\documentclass[amsmath,amssymb,prc,final,superscriptaddress,showpacs,twocolumn]{revtex4} 

\usepackage{bm,hyphenat,xspace} 
\usepackage{graphicx,epsfig}

\newcommand{\scl}{0.63}

\newcommand{\Ref}{Ref.}
\newcommand{\Refs}{Refs.}

\newcommand{\TR}{T^{(R)}(Z)}

\begin{document}

\title {New calculation schemes for proton-deuteron scattering including
the Coulomb interaction}

\author{A.~Deltuva} 
\email{deltuva@cii.fc.ul.pt}
\affiliation{Centro de F\'{\i}sica Nuclear da Universidade de Lisboa, 
P-1649-003 Lisboa, Portugal }

\author{A.~C.~Fonseca} 
\affiliation{Centro de F\'{\i}sica Nuclear da Universidade de Lisboa, 
P-1649-003 Lisboa, Portugal }

\author{P.~U.~Sauer}
\affiliation{Institut f\"ur Theoretische Physik,  Universit\"at Hannover,
  D-30167 Hannover, Germany}
\received{February 9, 2006}


\begin{abstract}
The Coulomb interaction between the protons is included in the
description of proton-deuteron scattering using the
screening and renormalization approach in the framework of
momentum-space integral equations. Two new calculational schemes are 
presented that confirm the reliability of the perturbative approach
for treating the screened Coulomb interaction in high partial waves,
used by us in earlier works.
\end{abstract}

 \maketitle


In \Refs~\cite{deltuva:05a,deltuva:05d}
we included the Coulomb interaction between the protons 
in the description of proton-deuteron $(pd)$ scattering and of 
three-nucleon electromagnetic (e.m.) reactions involving ${}^3\mathrm{He}$.
The description is based on the Alt-Grassberger-Sandhas (AGS) 
equations~\cite{alt:67a} in momentum space. The Coulomb potential
is screened, standard scattering theory is applicable,
and the resulting scattering amplitudes are corrected  
by the renormalization technique of \Refs~\cite{taylor:74a,alt:78a} 
to recover the unscreened limit.
The special choice of the screened Coulomb potential 
$w_R = w\; e^{-(r/R)^n}$,  $n=4$ being optimal,
approximates well the true Coulomb one $w$ for distances 
$r$ smaller than the screening radius $R$  
and simultaneously vanishes rapidly for $r>R$; 
rather modest values of $R$ are sufficient in order to 
obtain results that are well converged.
However, $R$ is considerably larger than the range of the strong interaction.
As a consequence, the calculation of the three-particle transition
operators for nuclear plus screened Coulomb potentials requires
the inclusion of partial waves with angular momentum much higher
than required for the hadronic potential alone.
In our previous calculations \cite{deltuva:05a,deltuva:05d}
the perturbation theory for high two-particle partial waves, developed in 
\Ref~\cite{deltuva:03b}, was used and found to be a very efficient and reliable
technical tool for treating the screened Coulomb interaction in those
high partial waves. The reliability was established by
the stability of the results when varying the dividing line between 
partial waves included exactly and perturbatively.
In this report we present two alternative
calculational schemes that solve $pd$ scattering equations 
without recourse to the perturbative approach of \Ref~\cite{deltuva:03b}.
This will enable us to demonstrate \emph{directly} the reliability 
of the method for the treatment of the screened Coulomb interaction in
high partial waves used in \Refs~\cite{deltuva:05a,deltuva:05d}.

The first calculational scheme, much like the perturbative approach of
\Refs~\cite{deltuva:05a,deltuva:05d}, is based on the
isospin formalism for three nucleons and on the symmetrized AGS equations
\begin{subequations} \label{eq:AGSsym}
  \begin{align}  \label{eq:UR}
    U^{(R)}(Z) = {} & P G_0^{-1}(Z) + P \TR G_0(Z) U^{(R)}(Z),
    \\ \nonumber
    U_0^{(R)}(Z) = {} & (1+P) G_0^{-1}(Z) \\  & + 
    (1+P) \TR G_0(Z) U^{(R)}(Z),  \label{eq:U0R}
  \end{align}
\end{subequations}
with $ G_0(Z)$ being the free resolvent, $P$ the sum of the two cyclic 
permutation operators, $\TR$ the two-particle transition matrix
derived from nuclear plus screened Coulomb potentials, and 
$U^{(R)}(Z)$  and $U_0^{(R)}(Z)$ the three-particle transition matrices
for elastic and breakup scattering; their dependence on the 
screening radius $R$ is notationally indicated.
The AGS equations are solved as they stand without any
perturbative feature. In the practical realization of the solution, 
i.e., in calculating the Neumann series
for the on-shell matrix elements of the operators 
$U^{(R)}(Z)$  and $ \TR G_0(Z) U^{(R)}(Z)$ 
and in summing by the Pad\'e method, the most time consuming part
is the action of the permutation operator $P$.
However, in two-particle partial waves with sufficiently high 
total angular momentum $I > I_N$ the hadronic interaction can safely 
be neglected and the two-particle transition matrix $\TR$
becomes the two-particle screened Coulomb transition matrix $T_{R}(Z)$,
which has a nonvanishing contribution only from the proton-proton $(pp)$ 
interaction. Using the isospin factors of  $T_{R}(Z)$ and of $P$
and including the three-nucleon total isospin $\mathcal{T} = \frac12$
and $\mathcal{T} = \frac32$ states, it can be shown that 
\begin{gather}
T_{R}(Z) G_0(Z) P T_{R}(Z) = 0,
\end{gather}
and that general combination of operators is the
basic computational building block when calculating the Neumann series
for $U^{(R)}(Z)$  and $ \TR G_0(Z) U^{(R)}(Z)$.
Thus, the contribution of the $P$ matrix elements between states with 
$I > I_N$ is zero and does not have to be calculated numerically.
Since typically $I_N$ is 4 to 6 and for the screened Coulomb interaction
states up to $I=12$ or even 14 have to be included, this yields a huge saving
of computer resources. Compared to our previous calculations based
on perturbation theory, the required computing time increases only
by a factor of 2 to 4; the corresponding calculations  can still 
be done on a PC, much like the ones of \Refs~\cite{deltuva:05a,deltuva:05d}.

The second calculational scheme considers protons and neutrons as
nonidentical particles. The identity of the two protons is taken into 
account by using properly antisymmetrized initial and final channel states.
The transition operators for elastic/rearrangement scattering and breakup 
are obtained from standard nonsymmetrized AGS equations
\begin{subequations}\label{eq:UbaT}
  \begin{align} \label{eq:Uba}
     U^{(R)}_{\beta \alpha}(Z) = {} & \bar{\delta}_{\beta \alpha} G_0^{-1}(Z)
     + \sum_{\sigma} \bar{\delta}_{\beta \sigma} T^{(R)}_\sigma (Z) G_0(Z)
     U^{(R)}_{\sigma \alpha}(Z), \\
     U^{(R)}_{0 \alpha}(Z) = {} & G_0^{-1}(Z)
     + \sum_{\sigma}  T^{(R)}_\sigma (Z) G_0(Z) U^{(R)}_{\sigma \alpha}(Z),
  \end{align}
\end{subequations}
with  $\bar{\delta}_{\beta \alpha} = 1 - {\delta}_{\beta \alpha}$.
The AGS equations are solved as they stand without any perturbative feature.
Here the most time consuming part of calculations is the transformation
from the basis where particle $\sigma$ is a spectator
to the basis where particle $\beta$ is the spectator, which is closely related
to the permutation operator $P$ of the first approach.
However, out of the three two-nucleon transition matrices $ T^{(R)}_\sigma (Z)$
only the one corresponding to a neutron spectator and a $pp$ pair
has high angular momentum  $I > I_N$ components due to the screened Coulomb
interaction. Therefore in the  transformation of bases the high
partial waves with $I > I_N$ will only be coupled to low partial waves
with $I \leq I_N$. The efficiency of this calculational scheme is 
nearly the same as of the first one.

The first calculational scheme is extended to include the $\Delta$-isobar 
excitation \cite{deltuva:03c} as a mechanism for mediating an effective 
three-nucleon force. In principle, a corresponding extension could be done also
for the second scheme. However, we refrain from doing
so in this report: Here we are only interested in the technical proof that
perturbation theory is highly reliable; furthermore, for the observables
at the energies considered the $\Delta$-isobar effects are insignificant.
We therefore present only results derived from the purely
nucleonic charge-dependent (CD) Bonn potential \cite{machleidt:01a}.
Using the two new calculational schemes we recalculated 
results of \Refs~\cite{deltuva:05a,deltuva:05d}. In all cases
we found excellent agreement, thereby confirming the reliability 
of the perturbative approach of \Refs~\cite{deltuva:05a,deltuva:05d}
for the Coulomb interaction in high partial waves.  
The examples for elastic $pd$ scattering
at 3 MeV proton lab energy and $pd$ breakup at 130 MeV deuteron
lab energy are shown in Figs.~\ref{fig:elastic} and \ref{fig:breakup}.

\renewcommand{\scl}{0.62}
\begin{figure}[b]
\begin{center}
\includegraphics[scale=\scl]{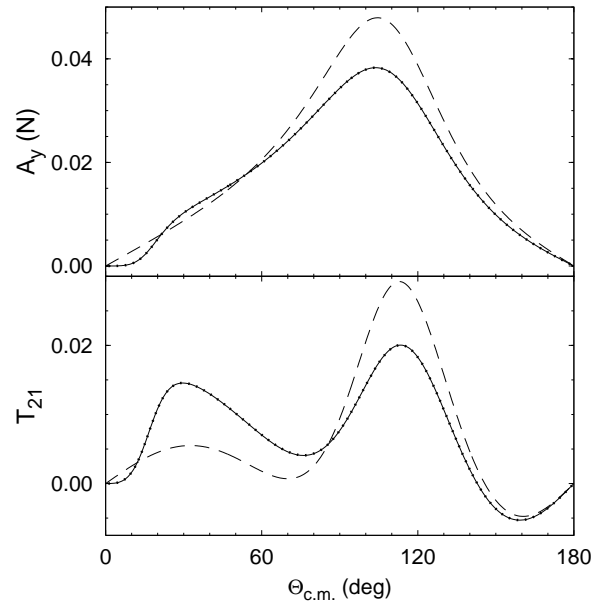}
\end{center}
\caption{\label{fig:elastic}
Nucleon analyzing power $A_y(N)$ and deuteron analyzing power $T_{21}$
for $pd$ elastic scattering at 3 MeV proton lab energy as functions of 
the c.m. scattering angle. The results of the two new calculational 
schemes without perturbation theory are graphically 
indistinguishable and are therefore represented by
one curve only, the thin solid one; they are compared with the results 
obtained  using perturbative approach of \Refs~\cite{deltuva:05a,deltuva:05d}
(thick dotted curves). Results without Coulomb (dashed curves)
are given as reference for the size of the Coulomb effect.}
\end{figure}

\begin{figure}[t]
\begin{center}
\includegraphics[scale=\scl]{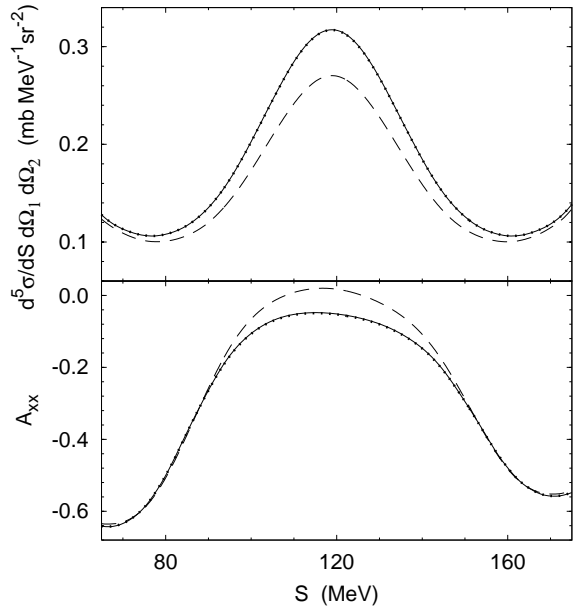}
\end{center}
\caption{\label{fig:breakup}
Differential cross section and deuteron analyzing power $A_{xx}$
for $pd$ breakup at 130~MeV deuteron lab energy in the
kinematical configuration $(15^{\circ},15^{\circ},160^{\circ})$ 
are shown as functions of the arclength $S$ along the kinematical curve.
Curves as in Fig.~\ref{fig:elastic}.}
\end{figure}

In summary,  we presented two new calculational schemes 
for treating the screened Coulomb interaction in high partial waves
without recourse to perturbative approach of \Ref~\cite{deltuva:03b}
used by us in previous works.
A perfect agreement between old and new results confirms
the reliability of the method used in \Refs~\cite{deltuva:05a,deltuva:05d}.

\vspace{1mm}

A.D. is supported by the FCT grant SFRH/BPD/14801/2003,
A.C.F. in part by the FCT grant POCTI/ISFL/2/275,
and P.U.S. in part by the DFG grant Sa 247/25.


\bibliographystyle{prsty}

\end{document}